\documentclass{ws-ijmpa}
\usepackage[super,compress]{cite}
\usepackage{graphicx}
\begin{document}
\markboth{Mario Campanelli}
{Instructions for Typing Manuscripts (Paper's Title)}

%
\catchline{}{}{}{}{}
%

\title{Forward jets and large rapidity gaps}

\author{Mario Campanelli\footnote{}}

\address{Department of Physics and Astronomy, University College London\\
Gower Street, WC1E 6BT London United Kingdom\\
mario.campanelli\@cern.ch}

\maketitle

\begin{abstract}
Hadronic jets are extremely abundant at the LHC, and testing QCD in various
corners of phase-space is important to understand backgrounds and some
specific signatures of new physics. In this article, various measurements
aiming at probing QCD in configurations where the theory modeling become 
challenging are presented. Azimuthal angle de-correlations are sensitive to
hard as well as soft QCD emission, and in most of the events jets are produced
in a back-to-back configuration. Events where jets have a large rapidity 
separation are also rare, and those without additional radiation between
the jets are exponentially suppressed. The modeling of radiation between
very forward and backward jets is complicated, and may require theoretical
tools different with respect to those normally used for central, high-pt
events. Observables can be created that are sensitive to all these effects,
like the study of azimuthal angle de-correlations between events where the
two leading jets have large rapidity separations. The two general-purpose
detectors of the LHC have measured these observables, and for some of them 
interesting deviations with respect to the most commonly used theoretical models
are observed.

\end{abstract}

\section{Introduction}	
Production of hadronic jet pairs is the most common high-momentum 
transfer process at the LHC, and has been widely studied, first with early
2010 data \cite{cms2010, atlas2010}, then with much larger datasets (e.g. 
\cite{cmsjetfull, atlasjetfull}). Dijet events are used to search
for new physics, in particular resonances \cite{cmssearches, atlassearches}
decaying into quark or gluon final states. Most of the standard measurements 
of dijet production are limited to jets of high transverse momentum in central 
pseudorapidities ($|\eta|< 3$). In this kinematic configuration, the
momentum fractions of the incoming partons are relatively large and of 
similar order of magnitude,
and in regions where strong constraints exist to the Parton Distribution 
Functions (PDF's) from deep-inelastic
scattering data. Also, for this high-$x$ kinematical configuration, it is 
expected that the standard DGLAP (Dokshitzer-Gribov-Lipatov-Altarelli-Parisi) 
\cite{dglap} evolution equations provide a good approximation of the 
underlying physics, and in
fact no sizeable deviations from the expected behaviour are observed.
More interesting tests of QCD can be performed in specific corners of 
phase-space, where the modeling of the underlying physics may be less obvious.
For instance, events with large azimuthal de-correlations are sensitive to
higher-order emission of hard gluons. Events with at least one jet in the
forward direction can arise from an imbalance of the momentum fraction of
the two partons, and therefore probe less-constraint regions of the PDFs.
Events where the two leading jets in the event have a large rapidity separation
may be better described by theoretical models involving multiple scales and
large logarithmic constributions. These configurations may require the use of
different evolution equations than DGLAP, like the approach from
Balitski-Fadin-Kuraev-Lipatov (BFKL) \cite{bfkl}, from 
Ciafaloni-Catani-Fiorani-Marchesini (CCFM) \cite{ccfm} or inspired by gluon 
saturation \cite{saturation}. \par
Evidence for deviations from the DGLAP description in systems of forward jets
has been searched for in various experiments and colliders before the LHC.
D0 measured forward-backward jets with rapidity separations up to 6 
\cite{d0dijets}, while both the ZEUS \cite{zeusforward} and H1 
\cite{h1forward} collaborations studied the systems with forward jets in the
final state.
Since these measurements did not give any compelling evidence for a 
breakdown of the validity of the DGLAP approach, it makes sense to 
pursue the study at the LHC, where the instrumentation of the detectors
in the very forward region allows probing even more extreme rapidity 
separations.
\par
In addition to the exchange of gluons varying QCD color, dijet events could 
also be produced by the exchange
of color-neutral gluon ladders. The probability of these kinds of processes 
is roughly independent on the rapidity separation between the dijets, while
the more common color-singlet exchange has an exponentially decreasing 
dependence on the rapidity separation. The consequence is that events with
large rapidity separation are proportionally more likely to be produced by
color singlets, a
configuration where no additional radiation between the jets is emitted.
The experimental study of events with jet veto, where only the events without 
hard emission in the rapidity region between the two leading dijets are
selected, can 
enhance the occurrence of interesting regions of phase space. Combination
between some of the techniques described above (like for instance the study
of azimuthal de-correlations for events with and without jet veto) can probe
even more specific regions of phase-space, highlighting the
potentially different ability to describe the data by the various theoretical
approaches.\par
In the following sections, several measurements are presented where QCD is
probed in specific corners of phase-space, with specific emphasis on 
measurements in the forward region of the detectors.

\section{Early azimuthal decorrelation measurements}
\begin{figure}[tbh]
\centerline{\includegraphics[width=0.45\linewidth]{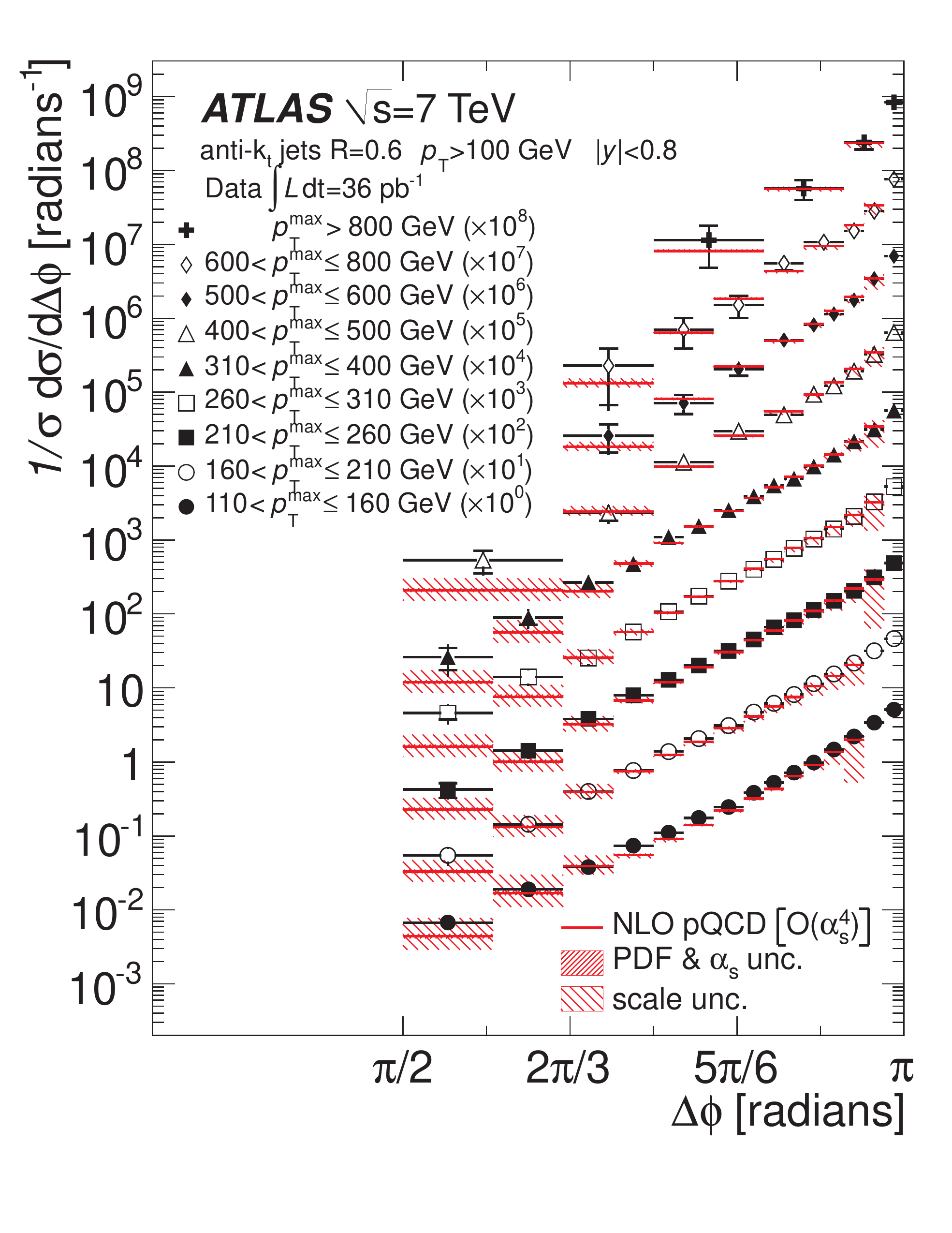}
\includegraphics[width=0.45\linewidth]{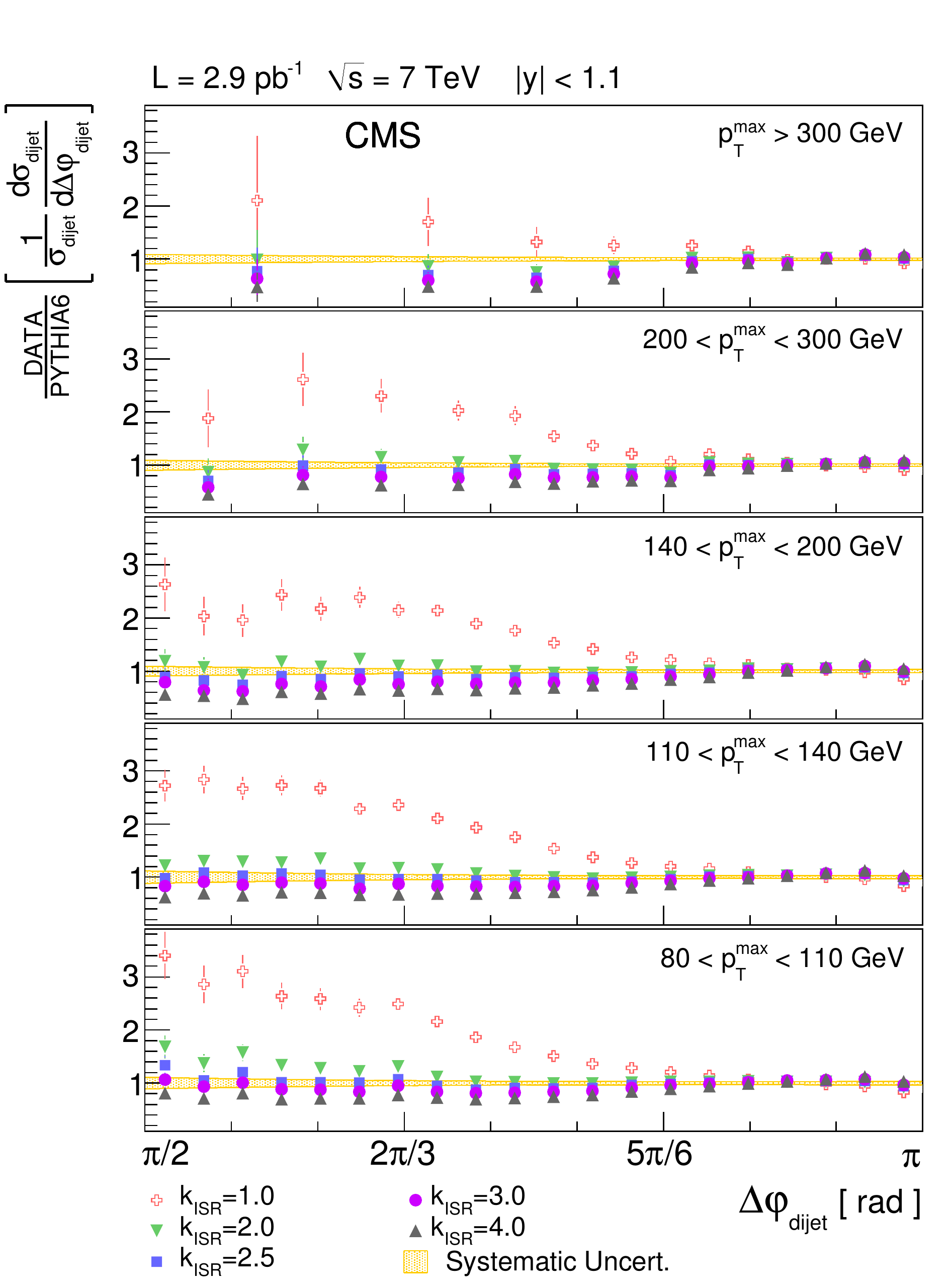}}
\caption{Azimuthal angle distribution in various ranges of the transverse
momentum of the leading jet. Left: data from ATLAS are compared to NLO QCD
predictions; right: the ratio between data and the D6T PYTHIA6 tune, where the
parameter governing ISR has been varied. \label{fig:azimuthal}}
\end{figure}

The first measurements to test QCD in difficult regions of phase-space of the
dijet system have been dijet azimuthal decorrelations. At Born level, the
outgoing partons in a 2$\rightarrow$ 2 interaction are produced exactly
back-to-back in the azimuthal plane, and with equal transverse momentum.
Hadronisation effects leading from partons to jets little change this picture,
so in the absence of extra radiation the angle between the two jets is supposed
to be very close to $\pi$, and indeed this is the case for the majority of
the events. Soft gluon radiation will contribute to small deviations with
respect to the back-to-back configuration, but it is only hard gluon radiation,
resulting in multijet final states, that can lead to significant deviations from
the back-to-back configuration. The measurement of azimuthal de-correlation is
therefore a test of QCD in various regimes, as well as a probe for various
Initial State Radiation (ISR) models, since hard ISR can boost the dijet system
in the transverse plane. Figure \ref{fig:azimuthal} shows the distribution of 
the azimuthal angle as measured by ATLAS using the data collected in 2010 
(left)\cite{atlas_azimuthal}, unfolded to particle level and compared to 
Next-to-Leading Order (NLO) QCD as computed by 
NLOJet++ \cite{nlojet}. A different approach is the one shown in the right plot
of that figure, where the ratio is shown of the aximuthal angle distribution
from data collected by CMS \cite{cms_azimuthal} to the predictions from D6T 
PYTHIA6\cite{pythia6} tunes.
In this case, the parameter responsible for
ISR [PARP(67)] has been varied over a wide range of values (including the
default of 2.5). It can be observed that the default value gives a very good
description of the data, while for extreme values like 1 the discrepancy with
data is large. Varying this parameter by $\pm 0.5$ gives variations at the 30\%
level, showing that the azimuthal decorrelation is sensitive to ISR, and can 
be used to tune this parameter.

\section{Cross section measurements in the forward region}
The early measurements of azimuthal decorrelations were performed on jets in the
central region, for reasons of trigger efficiency and uniformity of detector
response. Measuring the cross section for inclusive jets and dijets in the
forward region probes a region of the proton PDFs that is less constraint by
deep-inelastic scattering data, and also where perturbative QCD makes less
solid predictions. It was remarked, for instance, that for dijets with similar
transverse momenta, the cross section has a strong dependence on the choice
of renormalisation and factorisation scales, and that actually the usual
choice of setting these values to the transverse momentum of the leading jet
can lead to negative cross sections in some corners of the phase-space, such
as for large rapidity separations. In the ATLAS publication \cite{atlasdijets} 
the theory prediction had a choice
of scales that depends exponentially on $y^*$, half the absolute value of the
dijet rapidity separation:
\[\mu = p_T \exp(0.3 y^*) \]
\par
\begin{figure}[tbh]
\centerline{\includegraphics[width=0.45\linewidth]{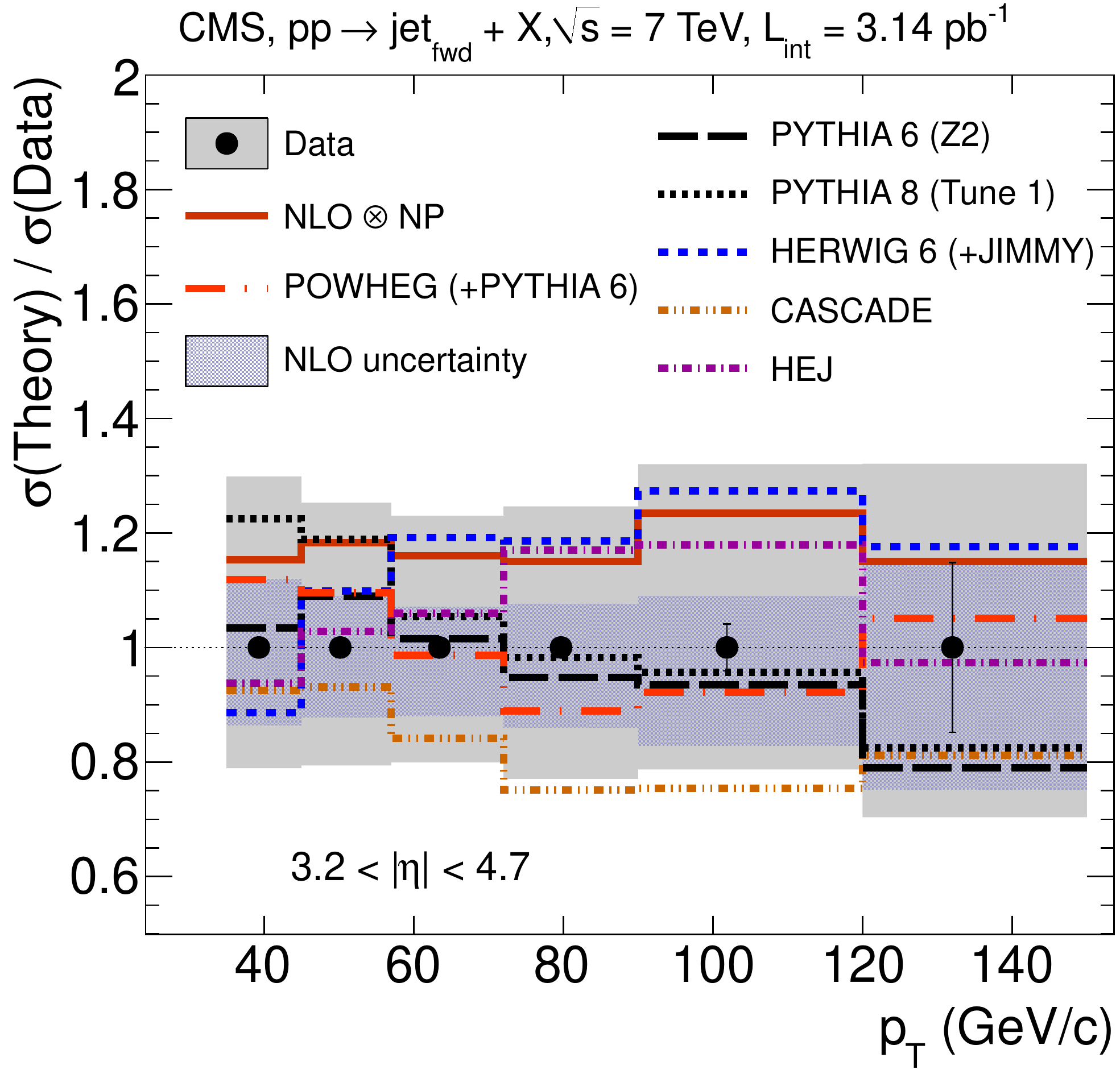}
\includegraphics[width=0.45\linewidth]{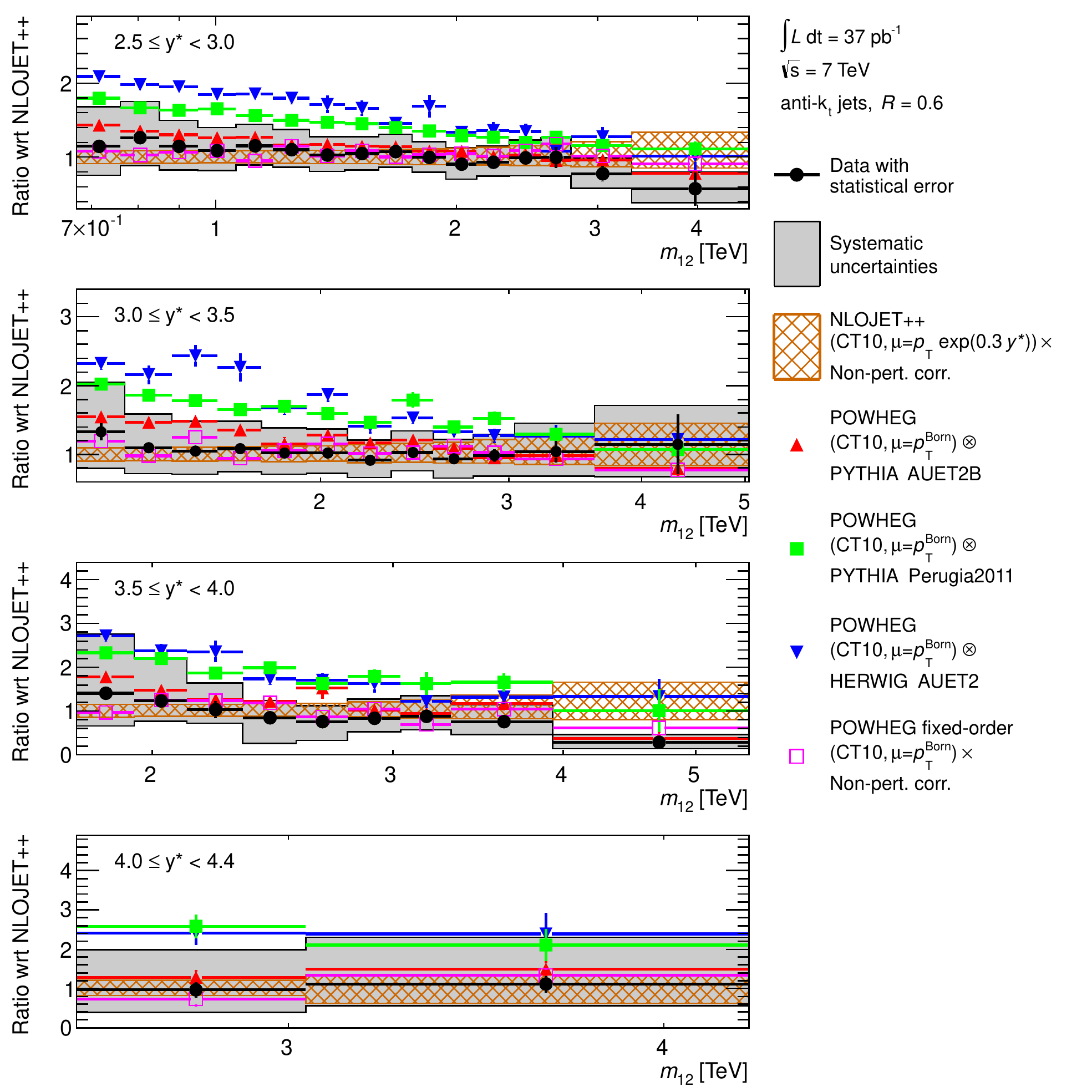}}
\caption{The ratio between measured dijet cross section in the forward
region for CMS (left) and ATLAS (right). Unfolded results from data are
compared to analytical NLO QCD calculations, as well as LO and NLO Monte Carlo
predictions.\label{fig:forwardxsec}}
\end{figure}

CMS measured the inclusive cross section for forward jets, and for dijets
where one of them is in the forward region, in Ref\cite{cmsfwdjets}. Unfolded
data are compared to a series of theoretical models, including NLO Monte Carlo
generators
and BFKL-inspired ones, in the rapidity range $3.2<|y|<4.7$. The ratio of many
theoretical models and data is shown in the left side of 
Fig. \ref{fig:forwardxsec}. Good agreement with all models is
present, within an uncertainty of 20\%, the same order of magnitude as
the theoretical differences.\par
ATLAS included the measurements of the forward region in the inclusive
and dijet cross section paper \cite{atlasdijets}. The right side of Fig.
\ref{fig:forwardxsec} shows the ratio of the ATLAS dijet cross section
measurement with respect to theory predictions from NLOJET++, compared to
the ratio of other theoretical models to the same denominator.
The cross section is measured as a function of the dijet invariant mass, in
various bins of the half rapidity difference $y^*$, of which the figure 
shows only the largest one.
Here the differences between the various models, as well as the systematic 
uncertainties for some points, can be up to factors of 2 or 3, being the 
largest at low transverse momenta and large rapidity separations.

\section{Dijets with large rapidity separations}
The kinematic regime where two jets are separated by a large rapidity separation
is where the standard modeling of QCD is expected to encounter more 
difficulties, and the approximations made in the DGLAP evolution to
break down. Alternative approaches to these evolution equation have been 
developed, both as effective theories \cite{lipatovefftheory} and as 
Monte Carlo generators inspired by CCFM evolution like CASCADE \cite{cascade} 
or BFKL like HEJ
\cite{hej}. The regime in which these effects become relevant is unclear, so
most of the experimental studies are performed as a function of the rapidity
separation between the two jets. \par
\begin{figure}[tbh]
\centerline{\includegraphics[width=0.45\linewidth]{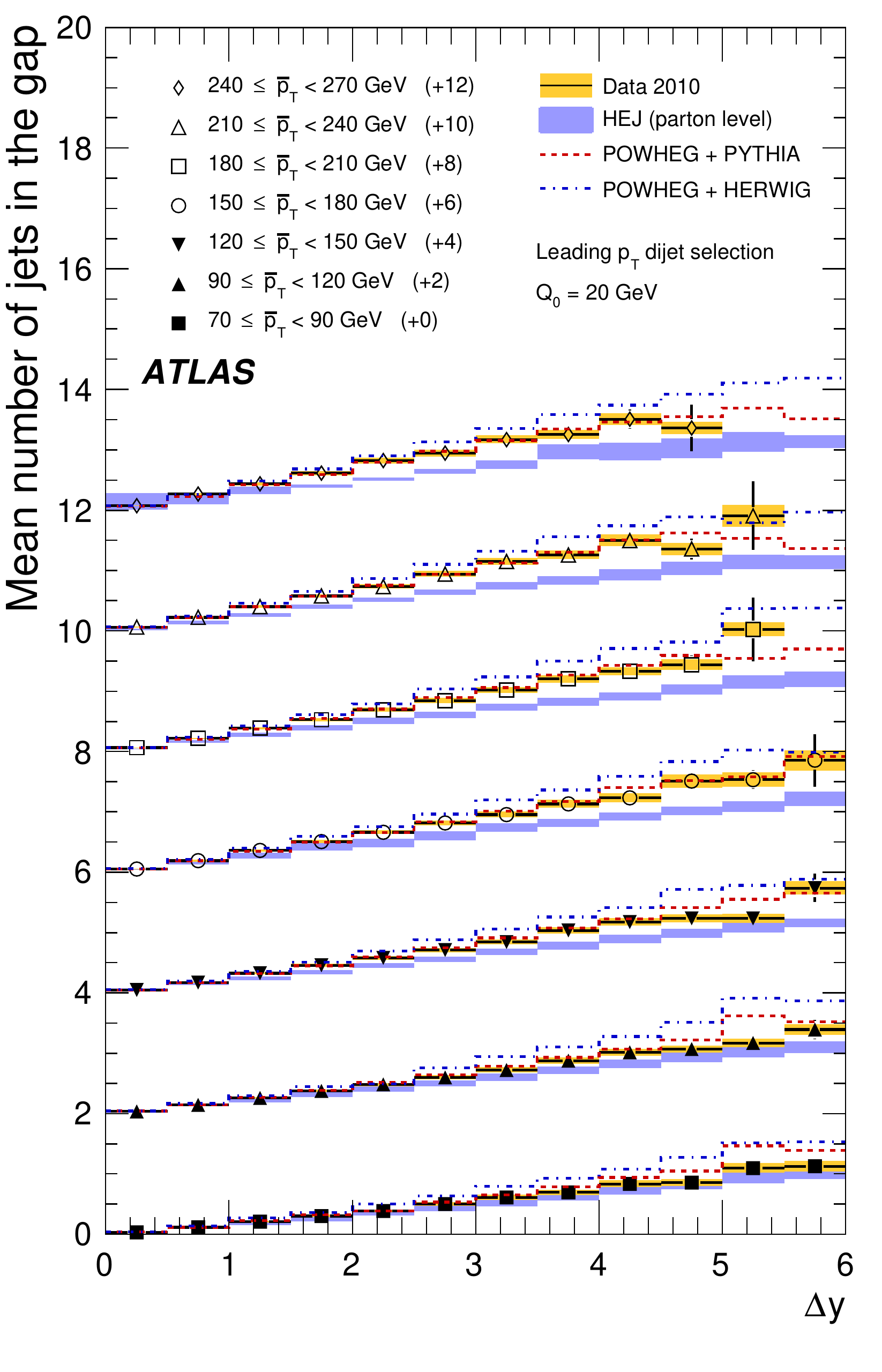}
\includegraphics[width=0.45\linewidth]{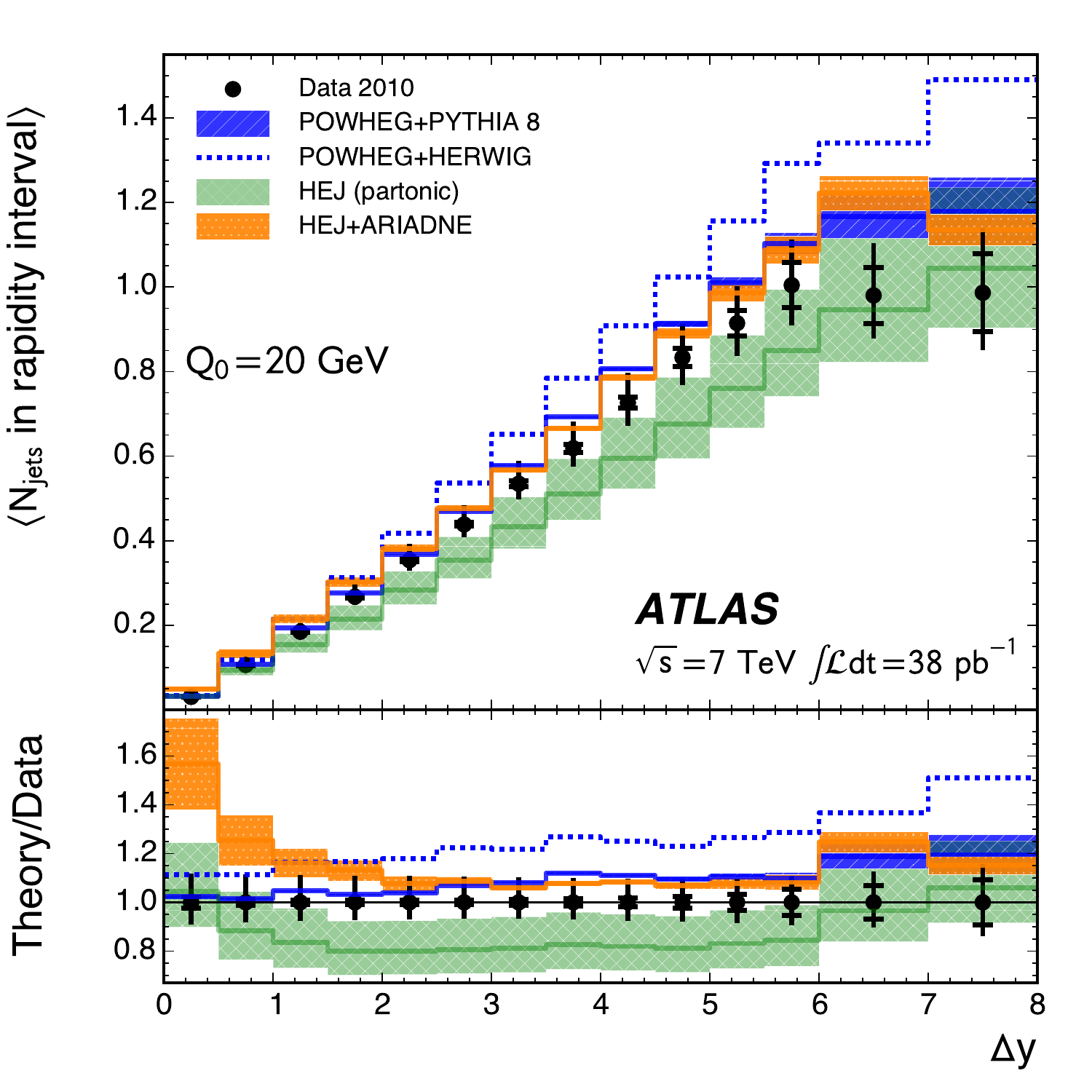}}
\caption{Mean number of jets in a rapidity gap, as a function of the rapidity
separation between the two leading jets in the event (left), and the separation
between the most forward and most backward jet (right plot).\label{fig:njetsvsrapidity}}
\end{figure}

ATLAS performed an explicit measurement of the number of additional jets in a 
dijet system, in the two cases when the boundary jets are defined as the two
leading jets of the events, and when they are the most forward and backward
of the event above a transverse momentum of 20 GeV (an approach inspired by
the studies of Mueller and Navelet \cite{muellernavelet}) 
\cite{atlasgappaper}. An additional requirement is that the 
average transverse momentum
of the two boundary jets should be larger than 50 GeV, to be in the 
fully-efficient region of the jet trigger. The measurements presented in Fig.
\ref{fig:njetsvsrapidity} are performed as a function of the dijet rapidity
separation, in bins of average transverse momentum of the two boundary jets.
The left plot shows the case when the boundary jets follow the Mueller-Navelet 
definition (the most forward and backward above a threshold). 
The theory comparison is made with
the NLO+PS generator POWHEG\cite{powheg}, interfaced with both 
PYTHIA and HERWIG\cite{herwig} for the
showering, and with HEJ interfaced with ARIADNE. Also in this case, the 
DGLAP-based generators, especially POWHEG interfaced with PYTHIA, perform very
well, giving an accurate description of the measured quantity over the whole
phase-space. HEJ tends to generally underestimate the radiation in the rapidity
interval, especially for the selection based on the leading jets in the event, 
and at large values of the average transverse momentum and rapidity separation.
The right plot, from a successive publication \cite{atlas2012}, shows the same
quantity in an extended range of $\Delta \eta$, in the case when the boundary
jets are the two leading jets of the event. The additional theory band when
the parton-level predictions from HEJ are coupled to a parton shower from 
ARIADNE \cite {ariadne} is present, showing a much better agreement with
data, even if the POWHEG+PYTHIA prediction is not significantly worse.
Also for this measurement, no advantage is seen in the use of specific 
BFKL-based models.

\section{Study of jet veto in dijet events}
\begin{figure}[tbh]
\centerline{\includegraphics[width=0.45\linewidth]{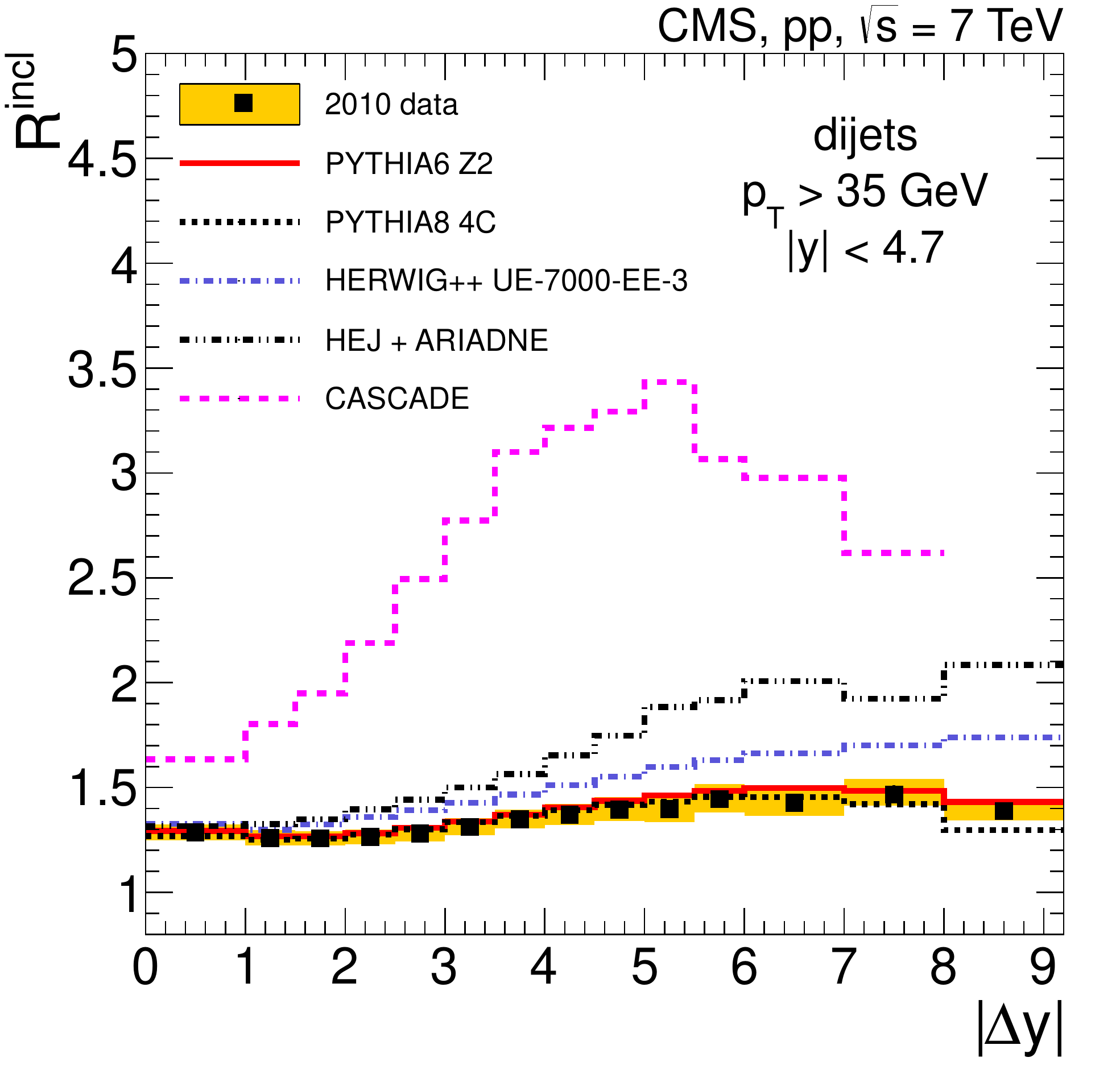}
\includegraphics[width=0.45\linewidth]{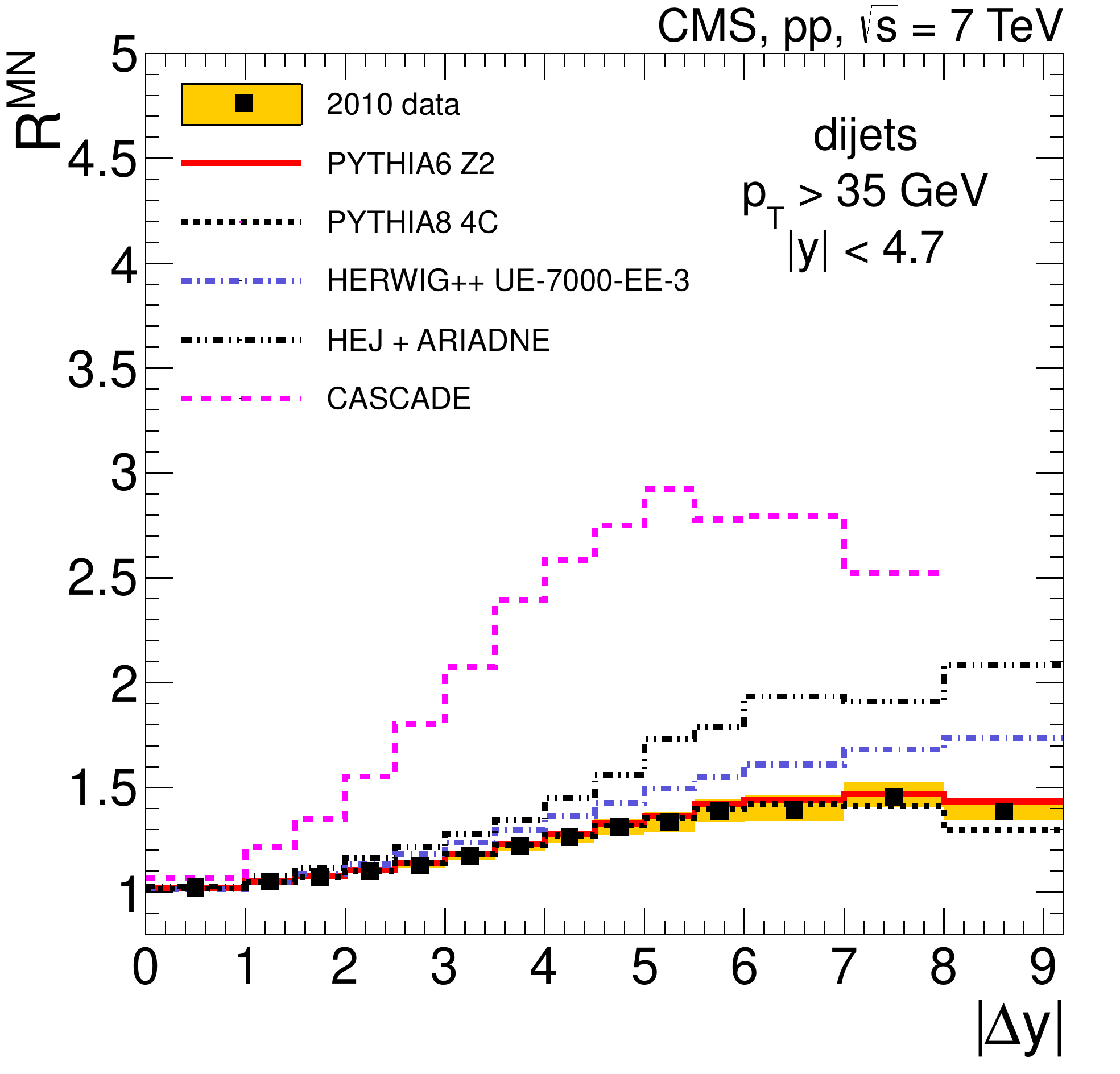}}
\caption{Left: ratio of the inclusive to exclusive dijet cross section
$R^{incl}$ as a function of the rapidity separation; right: ratio of the Muller-
Navelet to exclusive jet cross sections. Data with systematic error band (the statistical 
errors are smaller than the size of the symbol) is compared to several 
theoretical models. In addition to the DGLAP-based generators like PYTHIA6,
PYTHIA8 and HERWIG++, also BFKL-inspired ones like CASCADE and HEJ+ARIADNE are
shown.\label{fig:ratiosvsrapidity}}
\end{figure}

CMS measured \cite{cmsrapseparation} the ratios between inclusive and exclusive
dijet cross-sections $R^{\mathrm{incl}}= \sigma^{\mathrm{incl}}/\sigma^{\mathrm{excl}}$ and the ratio of the Mueller-Navelet jets to the exclusive ones
$R^{\mathrm{MN}}= \sigma^{\mathrm{MN}}/\sigma^{\mathrm{excl}}$ as a function of the dijet rapidity separation.
Events were considered if at least two jets with $p_T >$ 35 GeV and absolute
rapidity $|y| <$ 4.7 were present. Inclusive events have at least two jets passing these
criteria; exclusive events have exactly two jets of this kind. In the inclusive
case, each pairwise combination of jets is considered to calculate the rapidity
separation, so by construction the first ratio is always larger than one.
Mueller-Navelet jets are the most forward and most backward jet of the event; so
for the second ratio only one combination is taken, and by construction
$R^{MN} \leq R^{incl}$; however, also $R^{MN}$ is larger or equal than one, since,
even if only one combination is taken, the number of events with jets according
to the inclusive definition is larger than that of those with exclusive jets.
At extreme rapidity separations, the two ratios tend to converge, since it is
very rare to have more than one jet combination with very large $|\Delta y|$.
The results of this measurement are shown in Fig. \ref{fig:ratiosvsrapidity}
for the two ratios $R^{incl}$ and $R^{MN}$, respectively in the left and right 
plots. Data is compared to both DGLAP-inspired models like PYTHIA6, PYTHIA8
and HERWIG++ and to the generators CASCADE and HEJ 1.3.2, whose
parton-level jets coupled to the parton-shower model from ARIADNE 4.12.
The increase of the two ratios with rapidity separation follows the expected
opening of additional phase-space for additional parton radiation; as expected
the two quantities become similar at large $|\Delta y|$. The comparison with
theory shows that DGLAP-inspired generators (especially the two versions of 
PYTHIA) do a very good job at describing
data even in the region of large rapidity separations, where they were supposed
to perform badly; on the other hand, the two BFKL-inspired codes show large
discrepancies with data, for both ratios.\par

\begin{figure}[tbh]
\centerline{\includegraphics[width=0.45\linewidth]{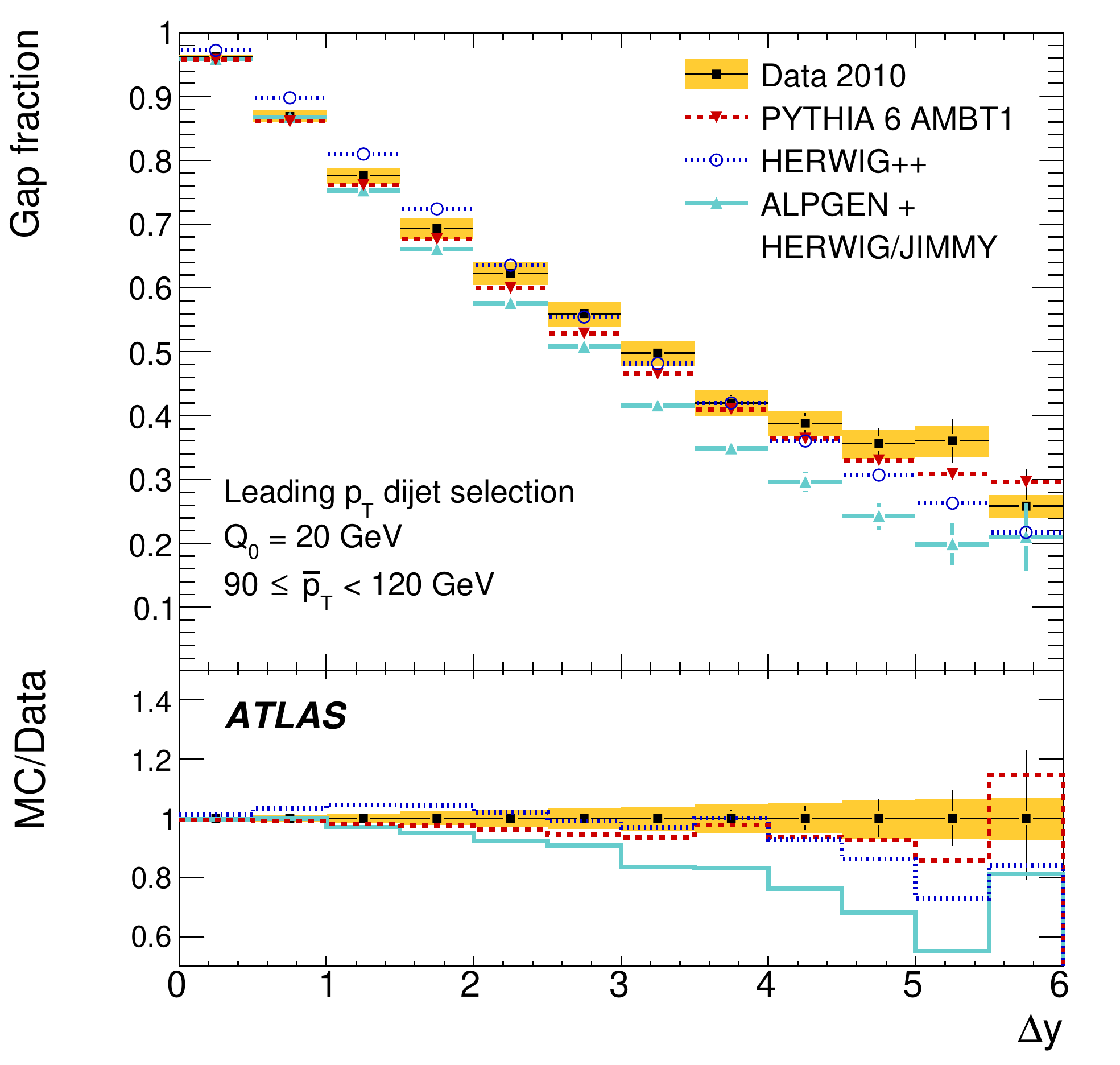}
\includegraphics[width=0.45\linewidth]{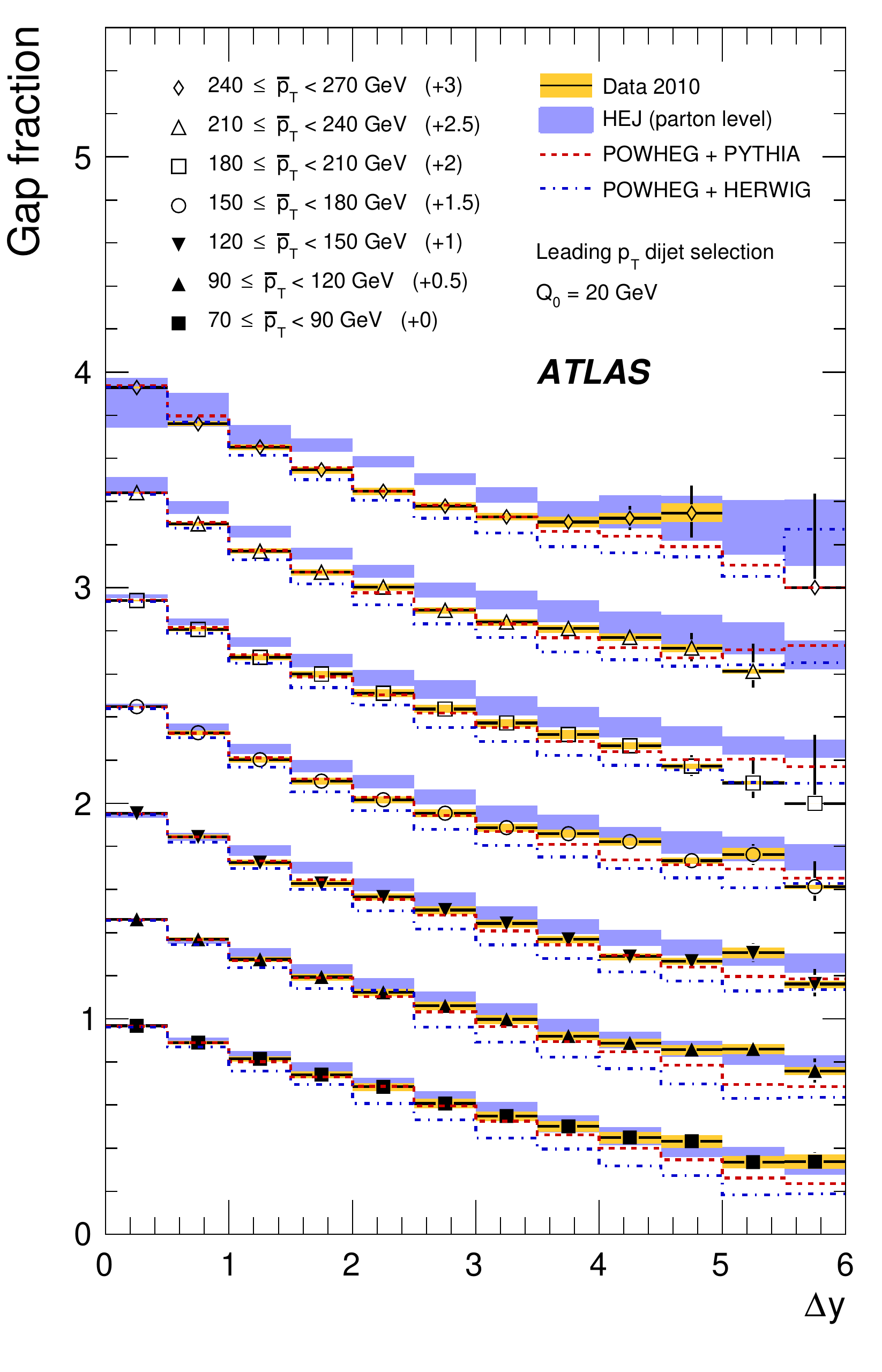}}
\caption{Gap fraction as a function of the rapidity separation between the two
leading jets in the event. Left: comparison of data to LO generators. Right: comparison with NLO and resummed generators.\label{fig:gapfraction}}
\end{figure}

ATLAS measured the ``gap fraction'', defined as the ratio between dijet
events without a third one in the rapidity interval between the two main
boundary jets and the total number of dijet events. As for the result on the
number of jets in the gap, two definitions of boundary jets were used: the two
leading jets in the event, and the most forward and backward above a given
transverse momentum threshold. The ratio taken using the second definition 
is equivalent to the
inverse of $R^{MN}$ from the CMS paper. For these two definitions of boundary
jets, ATLAS measures the gap fraction as a function of the rapidity difference
for various bins of average transverse momentum. Figure \ref{fig:gapfraction}
shows the gap fraction where the boundary jets are defined as the leading jets
of the event; on the left side, data is compared to leading-order 
codes like PYTHIA, HERWIG and SHERPA, while on the right side it is compared
to the NLO code POWHEG (coupled to parton shower by both PYTHIA and HERWIG), and to the resummed
approach of HEJ. We see that among LO generators, PYTHIA gives a reasonable
description of the data, while HERWIG and in particular SHERPA show increasing
deviations at the largest rapidity separations. The combination of POWHEG and
PYTHIA again gives the best description of data, with only small deviations
for large rapidity separations, at high or low values of the average boundary
jet values; on the other hand, HEJ overestimates the gap fraction in large 
regions of the phase space.\par

\begin{figure}[tbh]
\centerline{\includegraphics[width=0.45\linewidth]{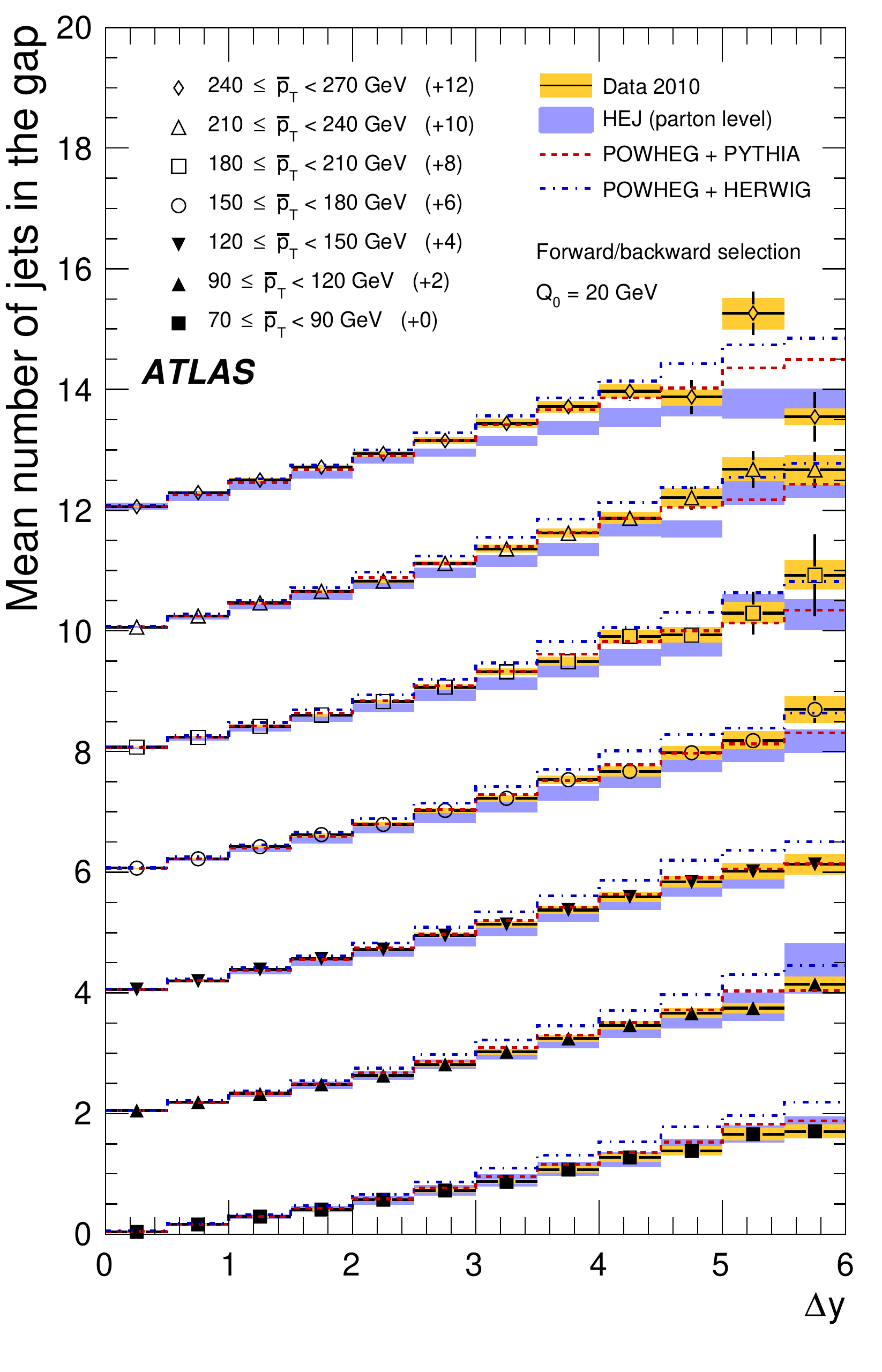}
\includegraphics[width=0.45\linewidth]{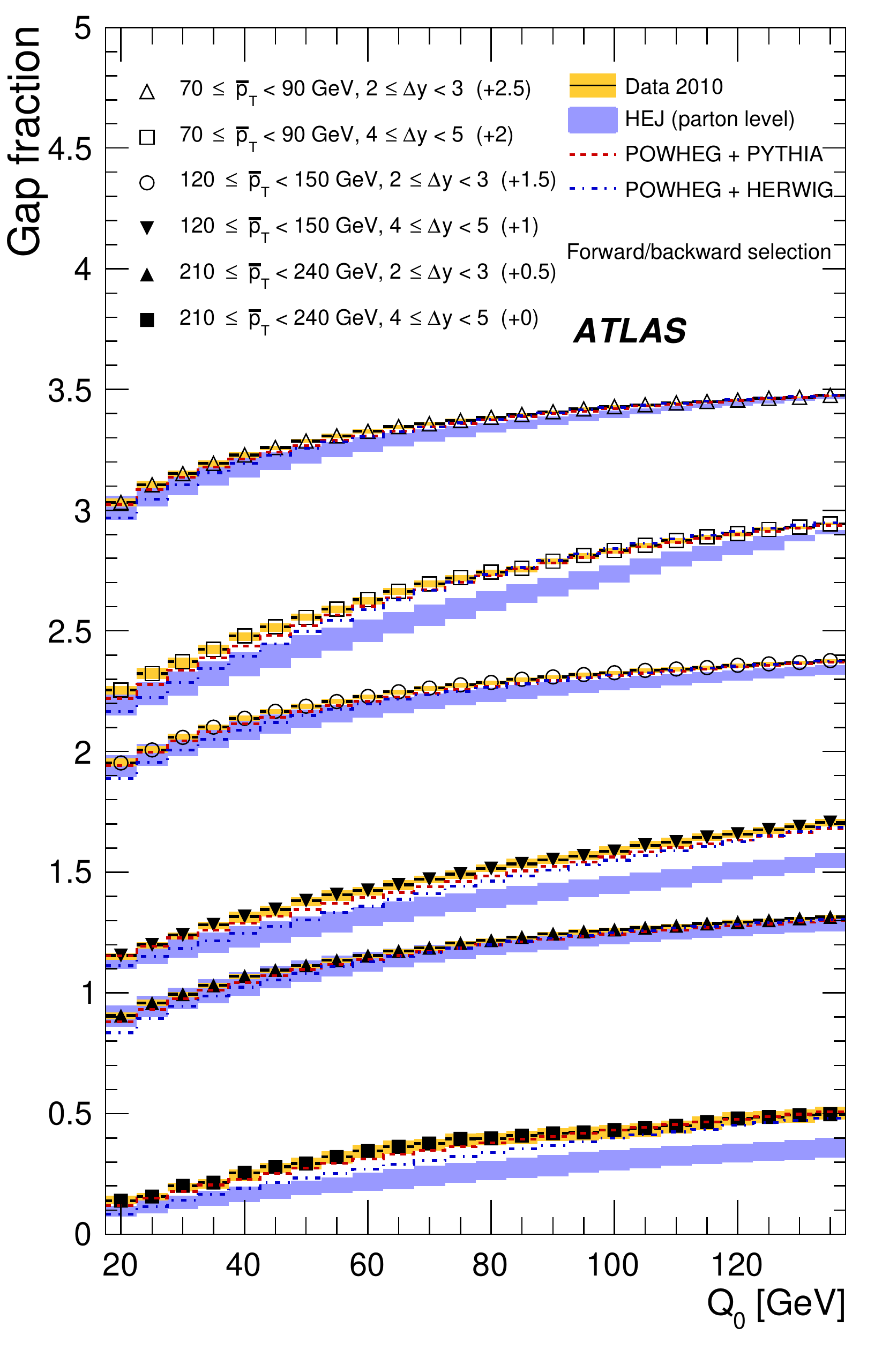}}
\caption{Gap fraction as a function of the rapidity separation between the most
forward and most backward jet in the event. Data is compared with NLO and
resummed generators. Left: gap fraction is plotted as a function of the rapidity
separation, keeping the threshold of the veto jet at 20 GeV; right: the gap
fraction for various combinations of average transverse momentum and rapidity
separation of the boundary jets is shown as a function of the threshold of the
veto jet.\label{fig:gapfractionMN}}
\end{figure}

Figure \ref{fig:gapfractionMN} shows instead the results obtained with the
Mueller-Navelet approach, when the two boundary jets are defined as the
most forward and most backward of the event. The left plot shows the gap 
fraction as a function of the dijet rapidity separation, and is therefore
directly comparable with the right side of Fig. \ref{fig:gapfraction},
but with a different boundary jet definition; the right side shows instead
the gap fraction as a function of the thresholds on the transverse momentum of
the veto jet. We see that, while in both cases the best agreement is again
reached by the POWHEG + PYTHIA combination, the agreement of HEJ is quite good
for the measurements where the veto threshold is fixed at 20 GeV, while it is
quite bad for larger values of the cut on the veto jet.

\section{Azimuthal decorrelations for forward dijets in the presence of jet veto}
Both azimuthal angle decorrelation and the internal jet veto probe the presence
of extra radiation in addition to (and within) the two boundary jets. To make more
specific tests of QCD, it makes sense to combine the two requirements for large
rapidity separations between the two boundary jets, measuring for instance 
the azimuthal angle as a function of $\Delta \eta$ for jets with and without
extra radiation in the rapidity gap.\par
\begin{figure}[tbh]
\centerline{\includegraphics[width=0.45\linewidth]{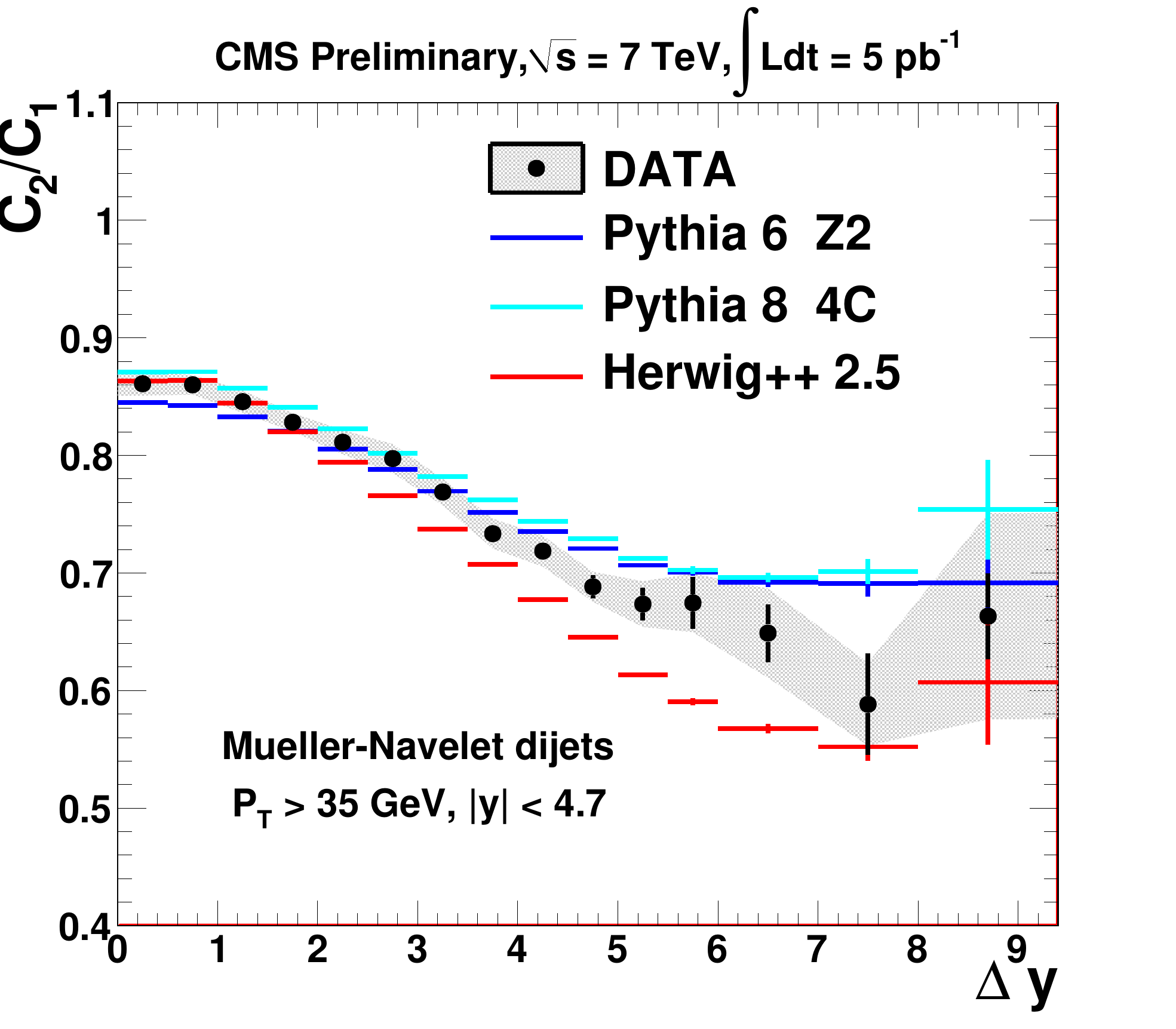}
\includegraphics[width=0.45\linewidth]{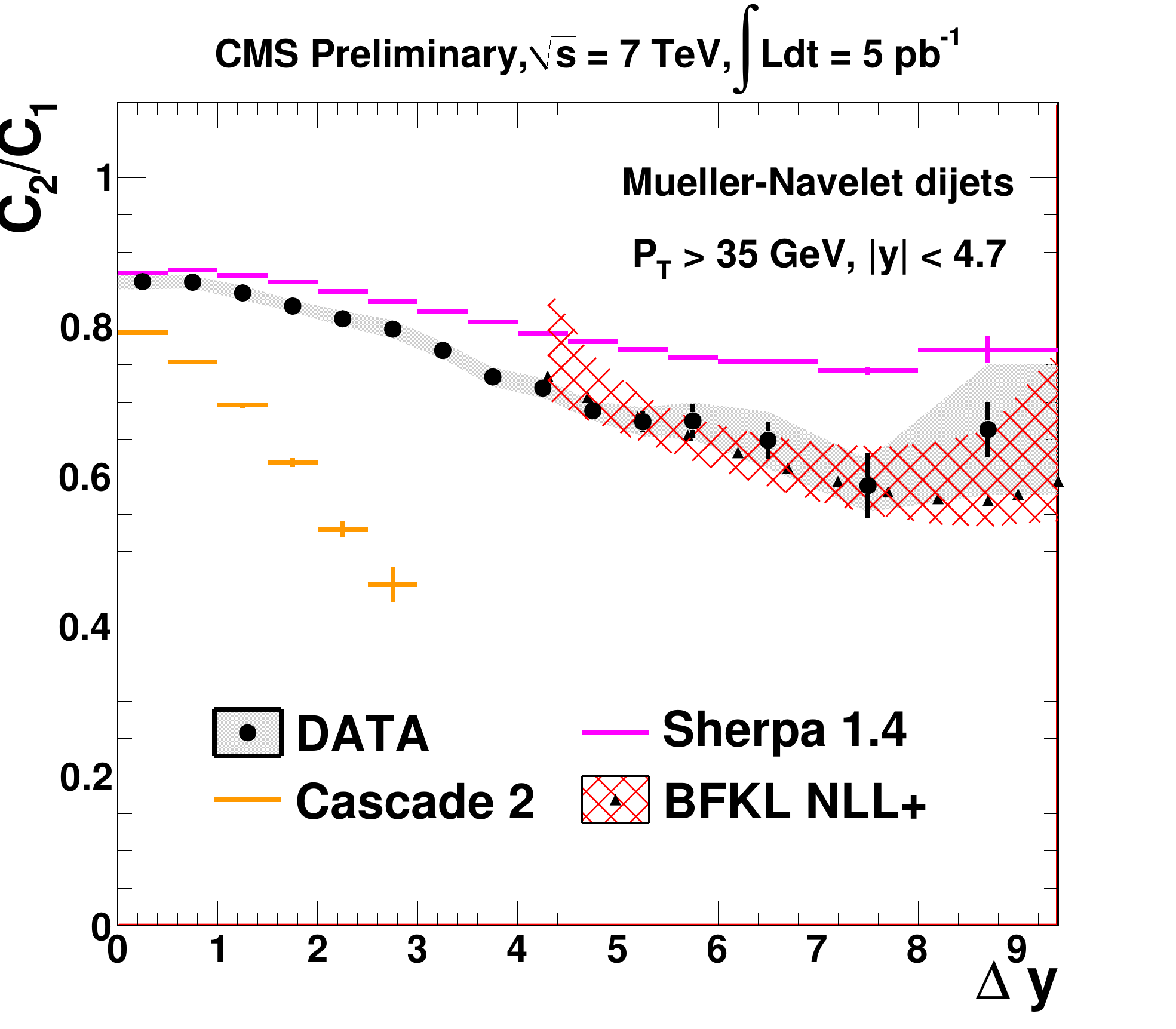}}
\caption{The ratio of the second and first Fourier coefficients $C_2/C_1$, as
a function of the rapidity separation between the jets. On the left, data is 
compared to LL DGLAP parton shower generators; on the left, to the 
multi-leg matrix element generator SHERPA, to the LL BFKL-inspired generator
CASCADE, and to a parton-level analytic NLL BFKL calculation.
\label{fig:c2c1cms}}
\end{figure}
\begin{figure}[tbh]
\centerline{\includegraphics[width=0.45\linewidth]{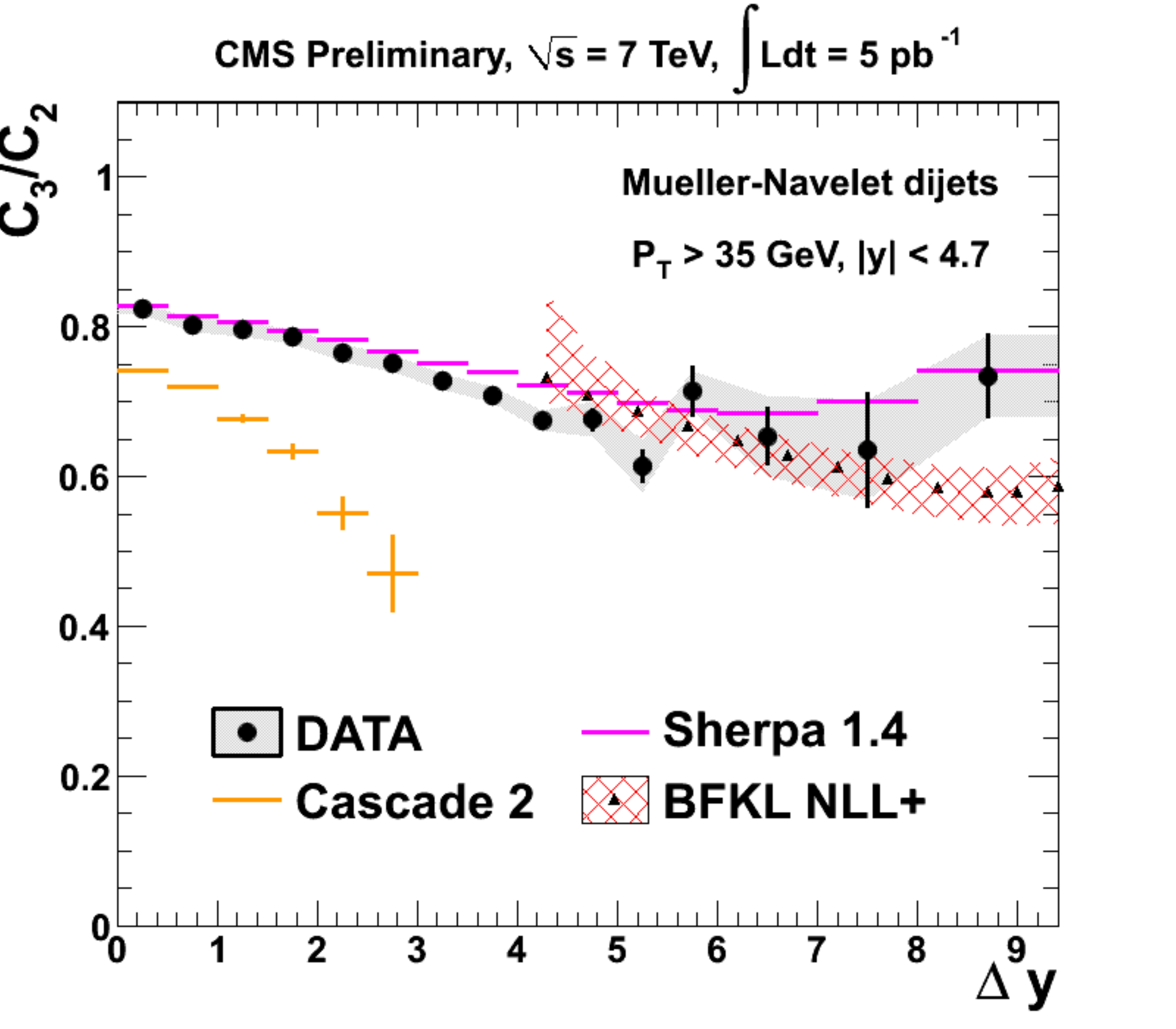}
\includegraphics[width=0.45\linewidth]{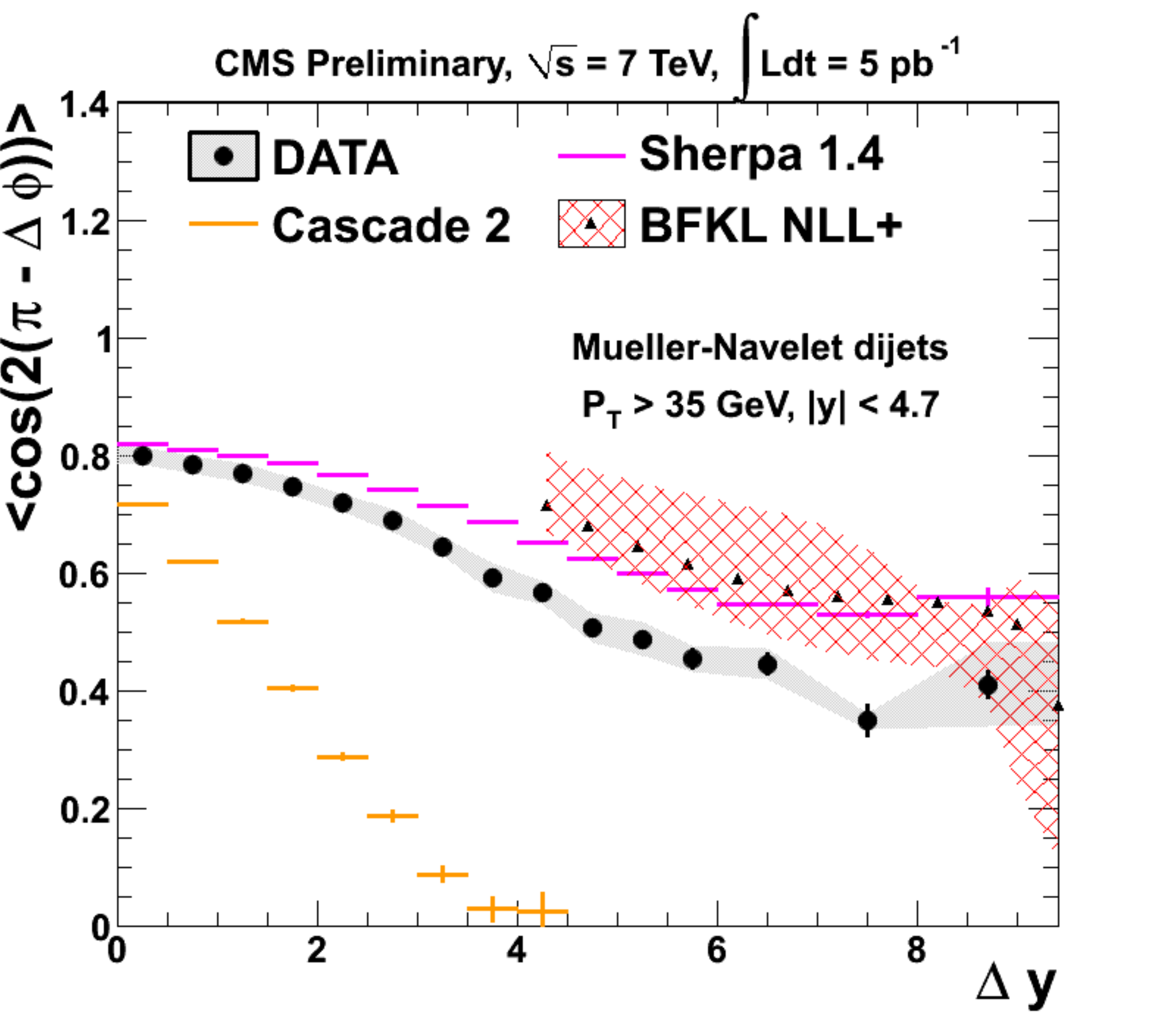}}
\caption{Left plot: the ratio of the third and second Fourier coefficients 
$C_3/C_2$, as a function of the rapidity separation between the jets, and 
compared to SHERPA, CASCADE, and BFKL NLL+. Right plot: the coefficient $C_2$,
compared to the same theoretical models.
\label{fig:coeffscms}}
\end{figure}

CMS produced two results where the azimuthal decorrelation between two jets
is measured as a function of their rapidity separation. 
In Ref. \cite{cmslargerap},
jets with rapidity separations up to 9.4 are considered. The normalised 
cross section for jets with transverse momentum above a given threshold 
$p_{Tmin}$ as a function of the azimuthal angle $\Delta \phi$ can be expanded 
in a Fourier series:
\[ \frac{1}{s}\frac{ds(\Delta y,p_{Tmin})}{d(\Delta\phi)}= \frac{1}{2p} [ 1 + 2 
\Sigma_{n=1}^{\inf} C_n(\Delta y,p_{Tmin}) \cos(n(p-\Delta\phi)) ]
\]
where the Fourier coefficients $C_n$ are equal to the average of the cosines
of the decorrelation angles multiplied by the order of the series $n$:
\[C_n(\Delta y,p_{Tmin}) = <\cos(n(p-\Delta\phi))>\]
At Born level, only two back-to-back jets are present, and at all orders
$C_n = 1$. Additional radiation leads to these coefficients becoming smaller
than one, and these decorrelations should increase with $\Delta y$, due to
the wider phase-space available to this extra radiation. The ratios of these
average cosines are particularly interesting \cite{sabiovera,angioni,ducloue} 
since on the one hand some experimental
uncertainties cancel out, on the other some DGLAP contributions are expected to
cancel \cite{angioni}, so BFKL-like effects are expected to be more visible.
Figure \ref{fig:c2c1cms} shows the ratio of the first two cosine coefficients,
as a function of the rapidity separation between the most forward and most
backward jet in the event. In the left plot data is compared to leading-order
generators, and some discrepancy is visible at large values of the rapidity
difference, even if the PYTHIA predictions are just on the upper side of the
1-sigma systematic error band. The right plot shows a comparison to a 
leading-logs BFKL-inspired generator (Cascade 2), to a matrix-element DGLAP
generator (SHERPA 1.4), and to a Next-to-Leading Logs (NLL) BFKL 
analysitcal calculation at parton
level \cite{ducloue}. In this specific ratio data is in very good agreement 
with this calculation, and this would be the first measurement so far showing
better agreement with a BFKL-inspired model than a DGLAP one; however the 
agreement with BFKL NLL+ gets worse in the ratio between 
coefficients 3 and 2, and in the values of the coefficients themselves, as
shown in figures \ref{fig:coeffscms}.

\par
\begin{figure}[tbh]
\centerline{\includegraphics[width=0.45\linewidth]{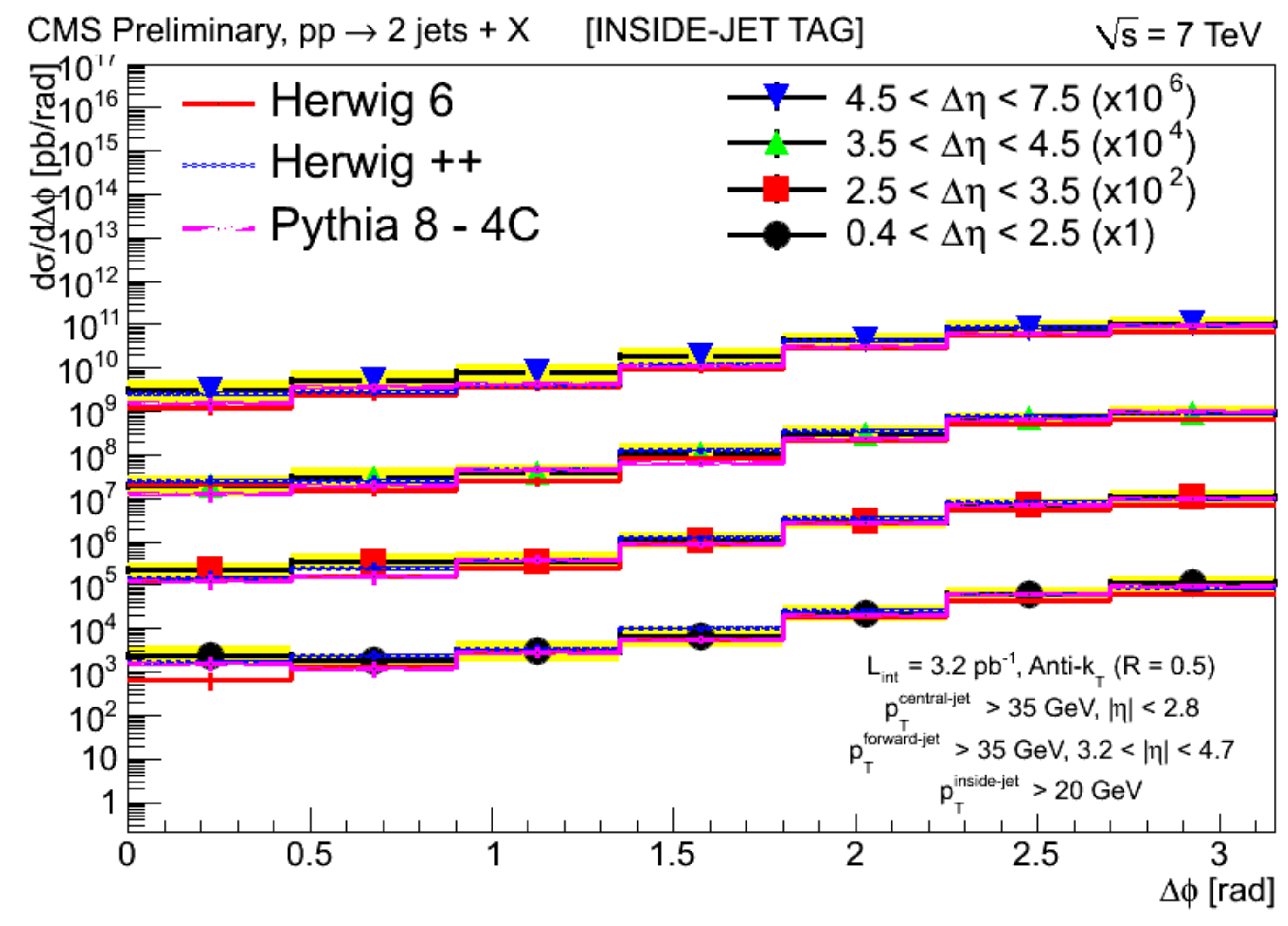}
\includegraphics[width=0.45\linewidth]{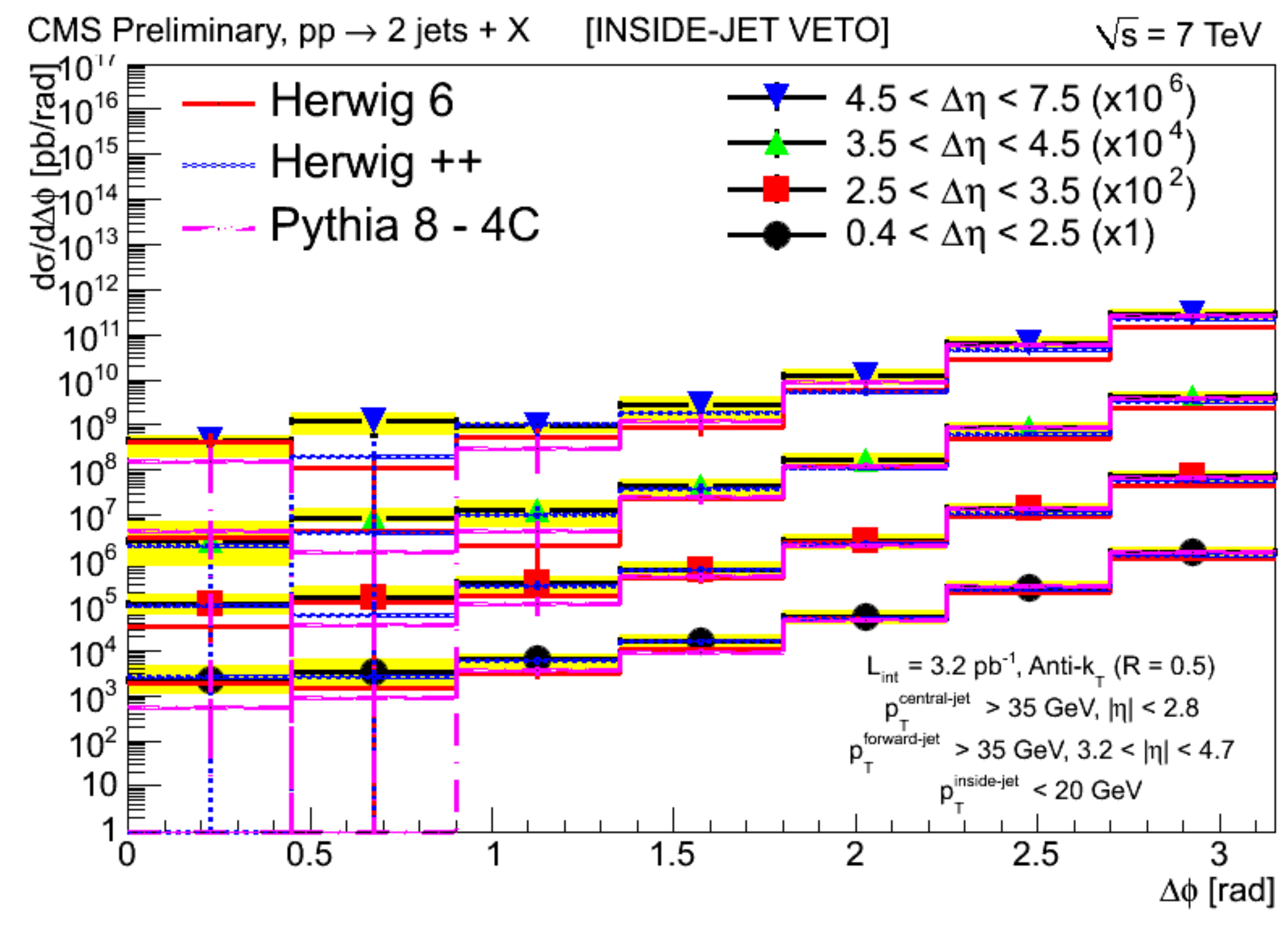}}
\caption{The dijet cross section as a function of the azimuthal angle between 
a central and a forward jet in various bins of their pseudorapidity separation
$\Delta \eta$. The left plot corresponds to events where an
additional jet is present between the two boundary ones, the right one to the
jet veto case. Data is compared to LO generators HERWIG 6, HERWIG++ and PYTHIA 
8.
\label{fig:cmsdphietaslices}}
\end{figure}

The second measurement \cite{cmsdphiveto} studies the azimuthal de-correlations
between a central and a forward jet separately for the cases where another jet
is present or for the case of a jet veto. Figure \ref{fig:cmsdphietaslices}
shows the dijet cross section as a function of $\Delta\Phi$ between a central
and a forward jet, for various bins of the pseudorapidity separation $\Delta
\eta$. Unfolded data is compared to the LO generators HERWIG 6, HERWIG++ and
PYTHIA 8. In general, good agreement is found, within systematic uncertainties
that can reach 50\% or more, while the difference between the models is about
half that value.

\par

Extending the study performed in Ref. \cite{atlasgappaper} to larger ranges 
in transverse momenta and
rapidity separations, ATLAS performed several measurements \cite{atlas2012}
both for all dijet events and for those with rapidity gaps.
Results are compared to the NLO generator POWHEG, where a matrix element 
calcultion is matched to the parton shower of PYTHIA8 and of HERWIG, as well
as to HEJ, at parton level and after showering by ARIADNE.
The measurement is performed in various bins of average tranverse momentum of
the leading jets, and of their rapidity separation.\par
\begin{figure}[tbh]
\centerline{\includegraphics[width=0.45\linewidth]{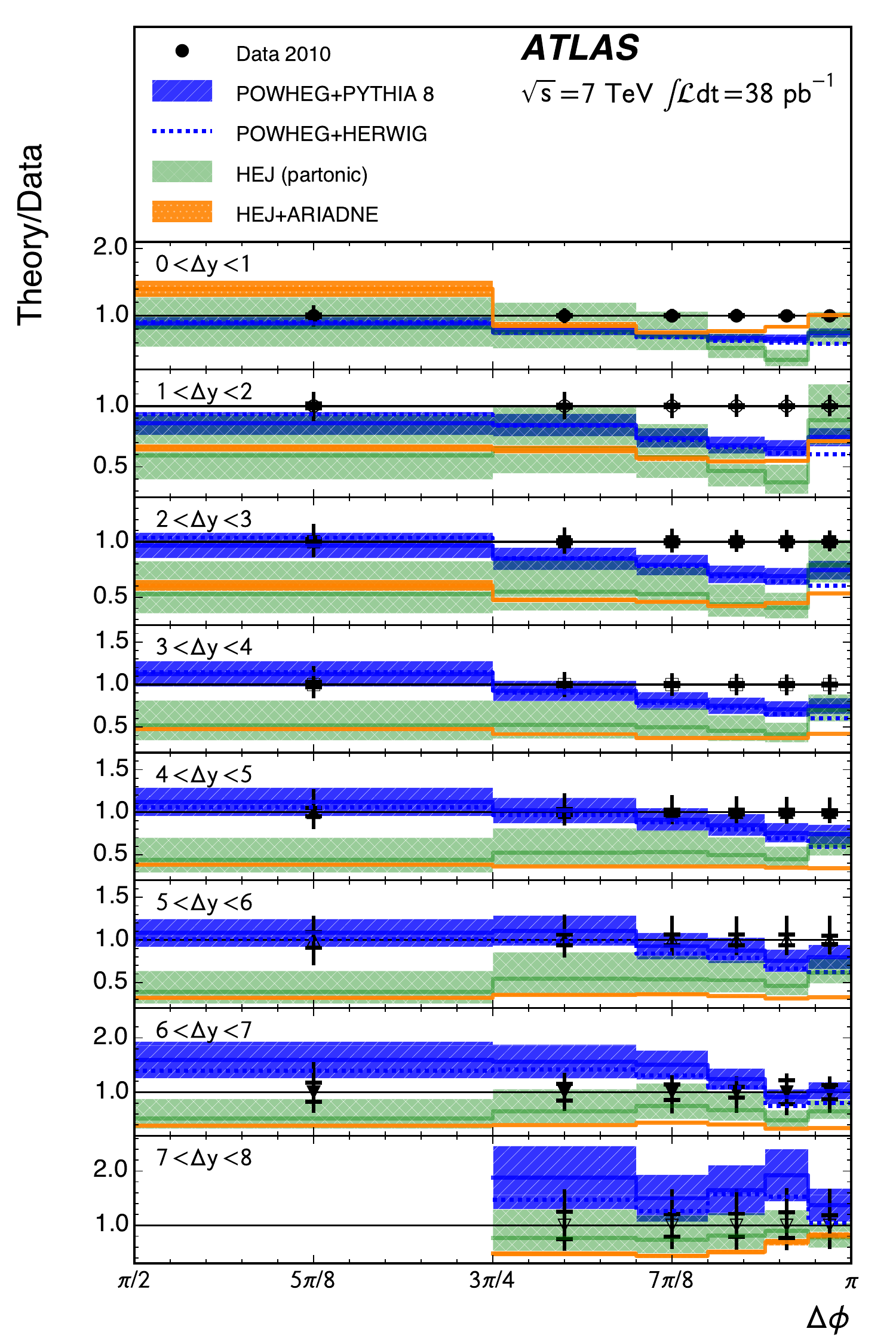}
\includegraphics[width=0.45\linewidth]{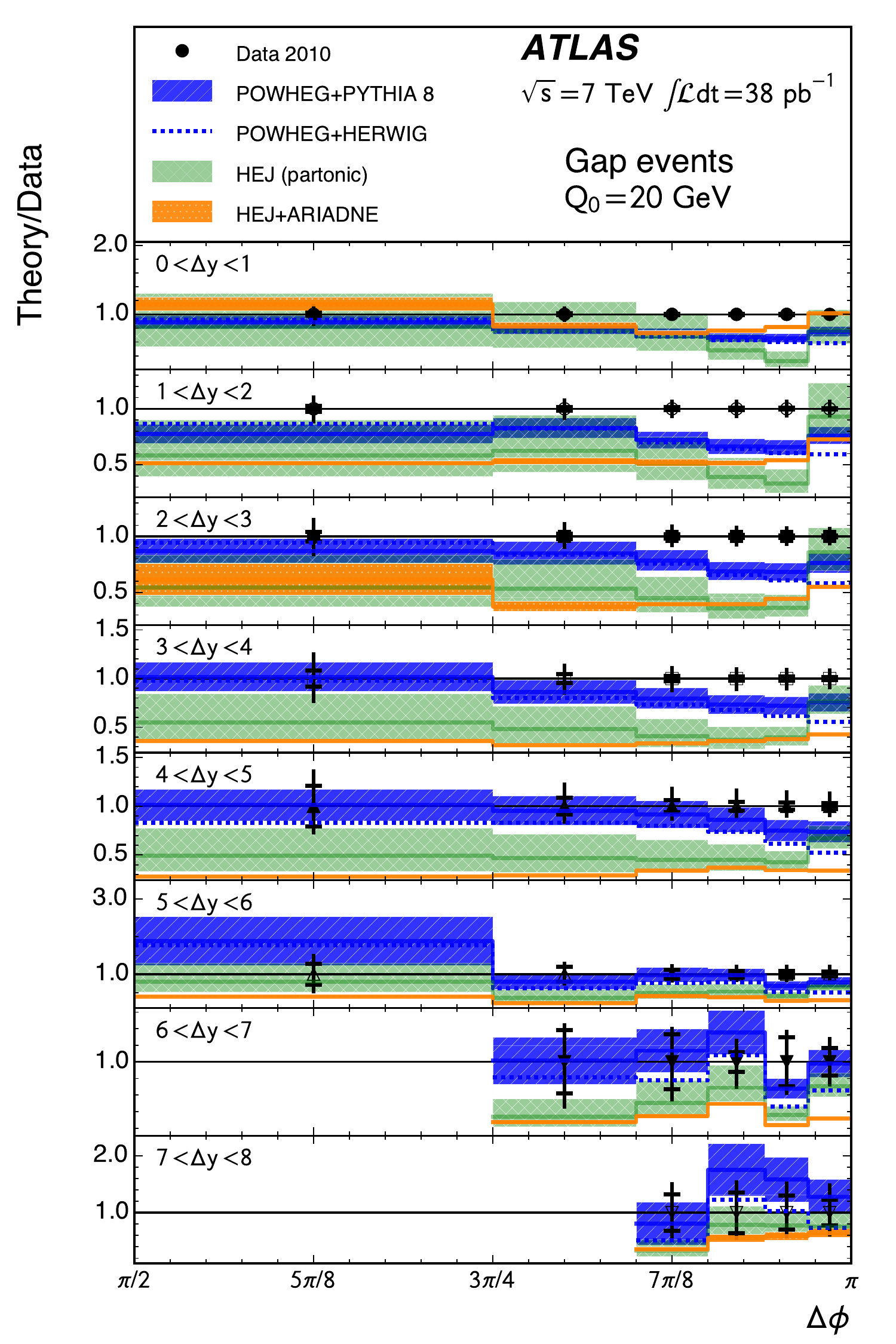}}
\caption{The dijet cross section as a function of the azimuthal angle
difference between the leading dijets in the event. The measurements
are presented as ratios between theory predictions and data, for all events
(left), and those without a jet with transverse momentum larger than 30 GeV
in the rapidity interval between them.\label{fig:atlasphidist}}
\end{figure}

Figure \ref{fig:atlasphidist} shows the ratio of the various theoretical
models to data, for various bins of rapidity separation between the two 
leading jets. The left plot is for all events, the right one is for events
with a rapidity gap, namely no additional jets above a transverse momentum 
$Q_0 = 30$ GeV in the rapidity range between the leading jets. As for the CMS 
measurement, systematic errors (added to the theoretical error bands) can be
large; however the difference between the various models can be larger than
these uncertainties. In general, apart from the smallest rapidity separation,
HEJ tends to underestimate the cross section, but has quite a similar slope
with respect to data, while POWHEG, with both showers, albeit being in general
in better agreement with data for small $\Delta\Phi$ values, tends as well to
underestimate the cross section as the azimuthal angle difference approaches 
$\pi$, so the ratio has a more marked slope.\par

\begin{figure}[tbh]
\centerline{\includegraphics[width=0.45\linewidth]{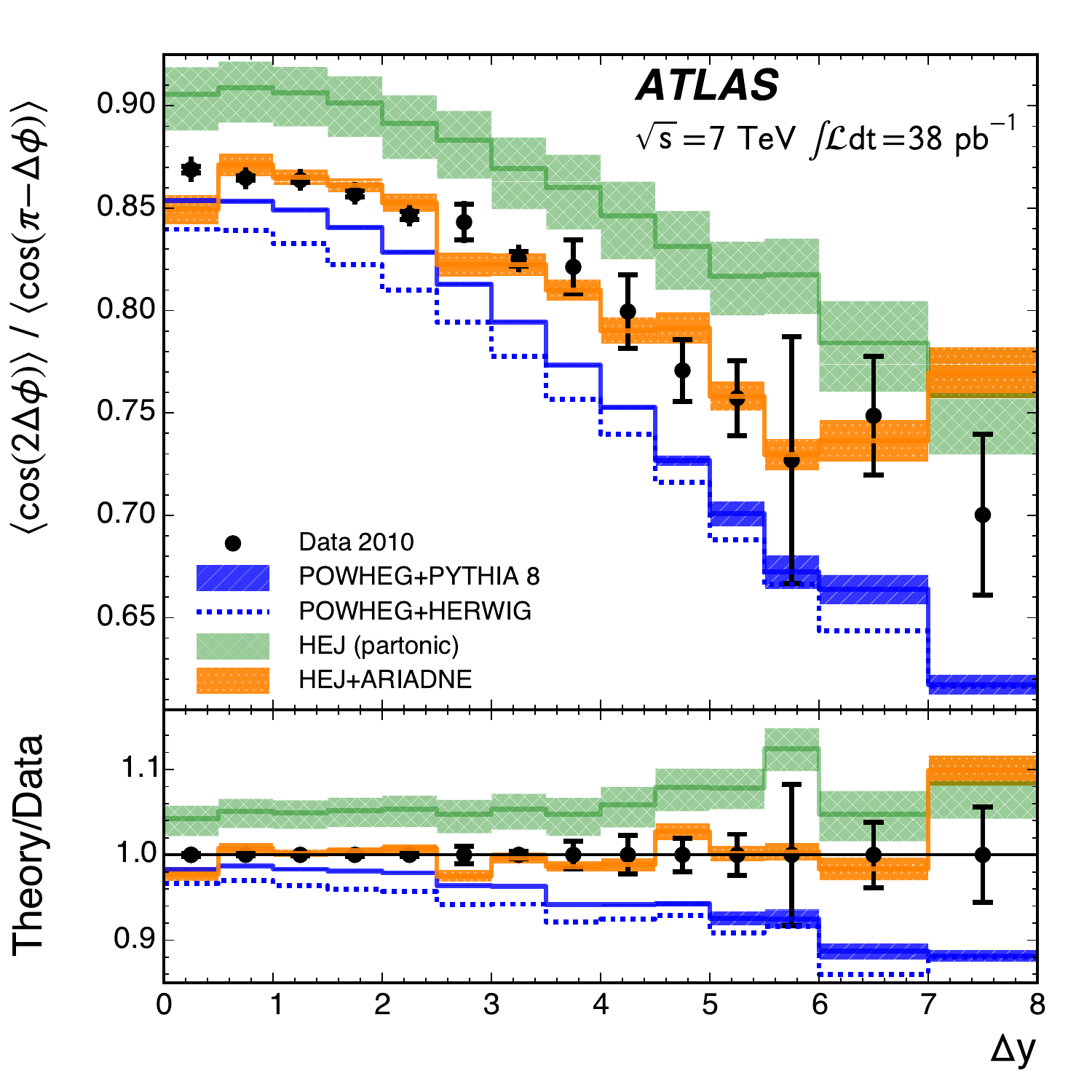}
\includegraphics[width=0.45\linewidth]{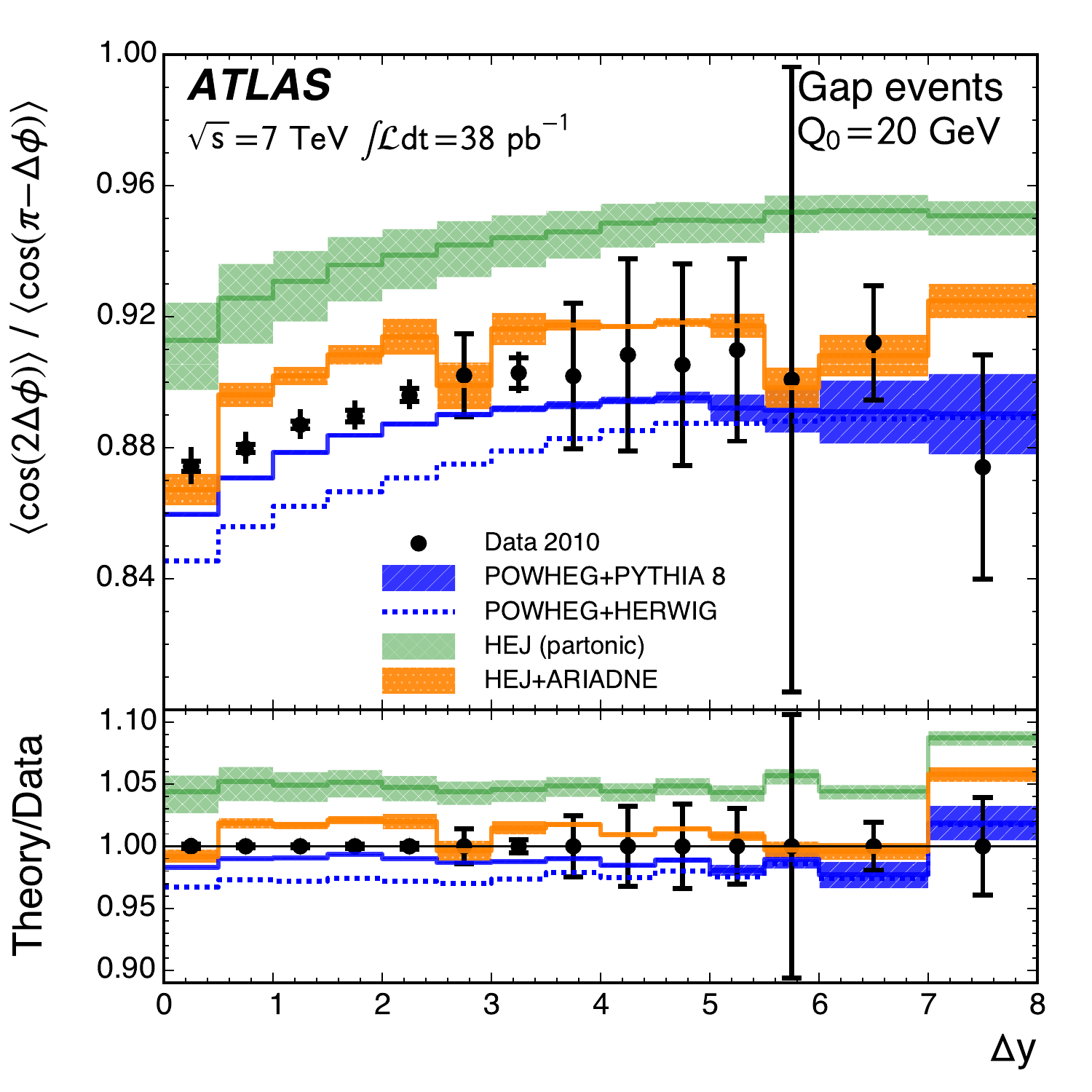}}
\caption{The ratio of the Fourier coefficients $C_2/C_1$, as a function of
the rapidity separation $\Delta y$ between the leading dijets in the event.
Left plot is for all events, right one is only for the events with a rapidity
gap, i.e. without another jet with transverse momentum larger than 30 GeV
in the rapidity interval between them.
\label{fig:atlasc2c1}}
\end{figure}

The plots in Fig. \ref{fig:atlasc2c1} show the ratio $C_2/C_1$ (the same as
on Fig. \ref{fig:c2c1cms}, but for a different boundary jet definition) as
a function of the rapidity separation between the jets, for all events (left)
and for those with a rapidity gap (right).
As for the CMS case, the ratio between the Fourier coefficients is found to be
a powerful variable to separate the various theoretical models, and a
case when data show significant disagreements with the DGLAP approach.
In particular the left plot, showing all events, seems to indicate a better
agreement with HEJ than with POWHEG, while this is not so clear in the case
(right plot) when a rapidity gap is requested. 

\section{Conclusions}
The two general-purpose experiments at the LHC have performed several 
measurements in particular corners of the jet production phase-space, 
to test QCD predictions in
regions where the simple predictions are expected to badly describe data.
Among them, the regime where jets are separated by large rapidity separations, 
and therefore produced in the forward regions, is expected to be better 
described by
alternative evolution equations, and has been actively investigated.
However, in the many measurements presented NLO QCD predictions, formulated within
the standard DGLAP framework, still show an amazingly good agreement with data.
Only very specific measurements, like the ratios of the coefficients of the
Fourier expansion of the difference of the azimuthal angle, start to show 
discrepancies, and possibly a better agreement
with the most recent BFKL-inspired predictions; but further measurements,
and better suited observables, are needed to state these conclusions in a more
definite way.

\end{document}